\documentclass[twocolumn,prl,superscriptaddress,floatfix,showpacs]{revtex4}

\usepackage{graphicx,latexsym}

\newcommand{\ctd}{\ensuremath{C_{\text{2D}}}}
\newcommand{\ceff}{\ensuremath{C_{\text{eff}}}}
\newcommand{\co}{\ensuremath{C_{1}}}
\newcommand{\ct}{\ensuremath{C_{2}}}
\newcommand{\csig}{\ensuremath{C_{\Sigma}}}
\newcommand{\cg}{\ensuremath{C_{g}}}

\newcommand{\ec}{\ensuremath{E_{c}}}

\newcommand{\rl}{\ensuremath{r_{\ell}}}
\newcommand{\cl}{\ensuremath{c_{\ell}}}
\newcommand{\iv}{$I$-$V$}
\newcommand{\rqu}{\ensuremath{R_{Q}}}
\newcommand{\kb}{\ensuremath{k_{B}}}
\newcommand{\rsq}{\ensuremath{R_{sq}}}
\newcommand{\ztd}{\ensuremath{Z_{\text{2D}}}}
\newcommand{\zl}{\ensuremath{Z_{\ell}}}
\newcommand{\zrc}{\ensuremath{Z_{RC}}}
\newcommand{\zt}{\ensuremath{Z_{t}}}
\newcommand{\grc}{\ensuremath{g_{rc}}}
\newcommand{\rqpc}{\ensuremath{R_{\text{QPC}}}}
\newcommand{\rstr}{\ensuremath{R_{\text{str}}}}
\newcommand{\cstr}{\ensuremath{C_{\text{str}}}}
\newcommand{\gset}{\ensuremath{G_{\text{SET}}}}
\newcommand{\gsc}{\ensuremath{\gset^{c}}}

\newcommand{\gtd}{\ensuremath{G_{\text{2D}}}}
\newcommand{\sch}{\ensuremath{\sigma_{ch}}}
\newcommand{\vg}{\ensuremath{V_{g}}}

\newcommand{\units}[1]{\ensuremath{\mathrm{#1}}}
\newcommand{\amount}[2]{\ensuremath{#1~\units{#2}}}
\newcommand{\e}[1]{\ensuremath{\times 10^{#1}}}
\newcommand{\etal}{\textit{et al.}}
\newcommand{\alxgas}{GaAs/Al$_{x}$Ga$_{1-x}$As}
\newcommand{\alx}{\ensuremath{{\rm Al/AlO}_{x}}}

\begin{document}

\title{Superconducting Single-Electron Transistor in a Locally Tunable
Electromagnetic Environment: Dissipation and Charge Fluctuations}
\author{W. Lu}
\affiliation{Department of Physics and Astronomy, Rice University,
Houston, Texas 77005}
\author{K. D. Maranowski}
\affiliation{Materials Department, University of California, Santa Barbara, 
California 93106}
\altaffiliation[Current address: ]{Cielo Communications, Inc., 325
Interlocken Parkway, Broomfield, CO 80021.}
\author{A. J. Rimberg}
\affiliation{Department of Physics and Astronomy, Rice University,
Houston, Texas 77005}
\affiliation{Department of Electrical and Computer Engineering, Rice University, 
Houston, Texas 77005}

\begin{abstract}
We have developed a novel system consisting of a superconducting
single-electron transistor (S-SET) coupled to a two-dimensional electron
gas (2DEG), for which the dissipation can be tuned in the immediate
vicinity of the S-SET\@. Within linear response, the S-SET conductance
varies nonmonotonically with increasing 2DEG impedance.  We find good
agreement between our experimental results and a model incorporating
electromagnetic fluctuations in both the S-SET leads and the 2DEG, as
well as low-frequency switching of the S-SET offset charge.
\end{abstract}


\maketitle

Electrical transport in nanoscale devices is strongly affected by the
electromagnetic properties of their environment.  This is particularly
true for superconducting systems, and has been a topic of considerable
interest lately: dissipation can drive a superconductor-insulator
quantum phase transition \cite{Rimberg:1997,Mason:1999,Penttila:1999},
and is also expected to affect the coherence time of qubits such as the
charge-based single Cooper pair box
\cite{Bouchiat:1998,Nakamura:1999,Makhlin:1999}.  Maintaining quantum
coherence in such devices long enough to allow many operations is
prerequisite for their use in quantum computation.  In this regard, the
closely related superconducting single electron transistor (S-SET) is an
excellent system for attaining a better understanding of the effects of
the electromagnetic environment on quantum coherence.

This approach was followed by the Berkeley group
\cite{Kycia:2001}, who fabricated an S-SET above a two-dimensional
electron gas (2DEG) in a \alxgas\ heterostructure.  Using a back gate to vary
the 2DEG sheet resistance \rsq\, they measured changes in the S-SET
conductance \gset\ as the dissipation was varied.  Building on earlier
theoretical work
\cite{Devoret:1990,Ingold:1992,Grabert:1998,Dittrich:1998,Ingold:1999},
Wilhelm, \etal\ predicted \cite{Wilhelm:2001} that within linear
response \gset\ would scale with the ground plane conductance
$\gtd=1/\rsq$ and temperature T as $\gtd^{\beta}/T^{\alpha}$ and that
the S-SET current $I$ at fixed bias voltage $V$ would vary
\textit{nonmonotonically} with \gtd, while switching from the non-linear
to linear regime.  These predictions were made within two models: in one
the environment was treated as a ground plane, while in the other it was
treated as an infinite $RC$ transmission line provided by the SET
leads.  While the Berkeley group did observe power law behavior, their
measured exponents were not in quantitative agreement with theory. 
Furthermore, the measured $\beta$ depended on $T$ and $\alpha$ on \gtd,
calling the scaling form into question.
\begin{figure}[h] \includegraphics[width=2.9in]{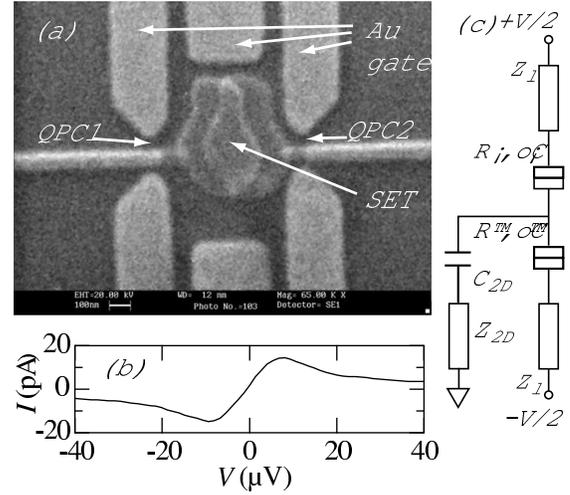}
\caption{\label{sample}(a) Electron micrograph showing the S-SET, gold
gates, and QPCs.  (b) S-SET \iv\ characteristics for $\vg=0$. The linear
regime extends to roughly $\pm\amount{8}{\mu V}$.  (c) Circuit diagram
of the S-SET, including its environment.  We allow both for lead
impedances \zl\ and a ground impedance \ztd\ coupled to the S-SET
through a capacitance \ctd.} \end{figure}

In this Letter, we report measurements on samples similar to those studied
by the Berkeley group; in our samples, however, we can modify \gtd\ in
the immediate vicinity of the S-SET while leaving the 2DEG beneath the
leads largely untouched.  In contrast to the predictions of Wilhelm,
\etal\ we observe a nonmonotonic dependence of \gset\ on \gtd\ entirely
within the linear regime.  We propose a model for the
environmental impedance of S-SET/2DEG systems that includes
electromagnetic fluctuations in both the 2DEG and leads, while treating
the latter as \textit{finite} $RC$ transmission lines.  We also allow
for relatively low-frequency switching of the charge state of the SET
island\cite{Eiles:1994}, which can affect the measured current. Within
this model, we find good agreement between our calculated and measured
results.

An electron micrograph of a typical sample is shown in
Fig.~\ref{sample}(a).  We begin with an \alxgas\ heterostructure grown
on a GaAs substrate using molecular beam epitaxy, consisting of the
following layers: \amount{1000}{nm} of GaAs, \amount{47}{nm} of
Al$_{0.3}$Ga$_{0.7}$As and \amount{5}{nm} of GaAs.  The
Al$_{0.3}$Ga$_{0.7}$As is delta-doped with Si \amount{22}{nm} from the
lower GaAs/Al$_{0.3}$Ga$_{0.7}$As interface, at which forms a
two-dimensional electron gas (2DEG) with $\rsq=\amount{20}{\Omega}$ and
sheet density $n_s=\amount{3.6\e{11}}{cm^{-2}}$.  On the sample surface
we use electron-beam lithography and shadow evaporation to fabricate an
\alx\ S-SET surrounded by six Au gates \cite{Lu:2000}. When no gate
voltage is applied and the 2DEG is unconfined, the measured \iv\
characteristics are linear over several microvolts, as shown in
Fig.~\ref{sample}(b). We can also apply a single gate voltage \vg\ to
any combination of Au gates, excluding the 2DEG beneath them. We focus
on two geometries: the ``pool,'' in which all six gates are energized,
and the ``stripe'' in which only the four exterior gates are. In both
cases, electrons immediately beneath the SET are coupled to ground by
quantum point contacts (QPCs) with conductances $1/\rqpc$ (assumed
equal) as low as 3 conductance quanta $G_{0}=e^2/h$.  In the stripe
geometry, the electrons can also move vertically through a resistance
\rstr\ to a large 2DEG reservoir that is coupled to ground through a
capacitance \cstr. As illustrated in Fig.~\ref{sample}(c),
electromagnetic fluctuations in the environment can couple to the
tunneling electrons in two ways: through the leads, which act as $RC$
transmission lines with impedance \zl\ for the relevant frequency range
\cite{Kycia:2001,Wilhelm:2001}, and through the capacitance \ctd\ to the
2DEG with impedance \ztd, which is related to \rqpc\ and (for the
stripe) \rstr. The model has been studied previously
\cite{Odintsov:1991,Ingold:1991} without considering particular
forms for \ztd\ and \zl.

\begin{figure}
\includegraphics[width=2.8in]{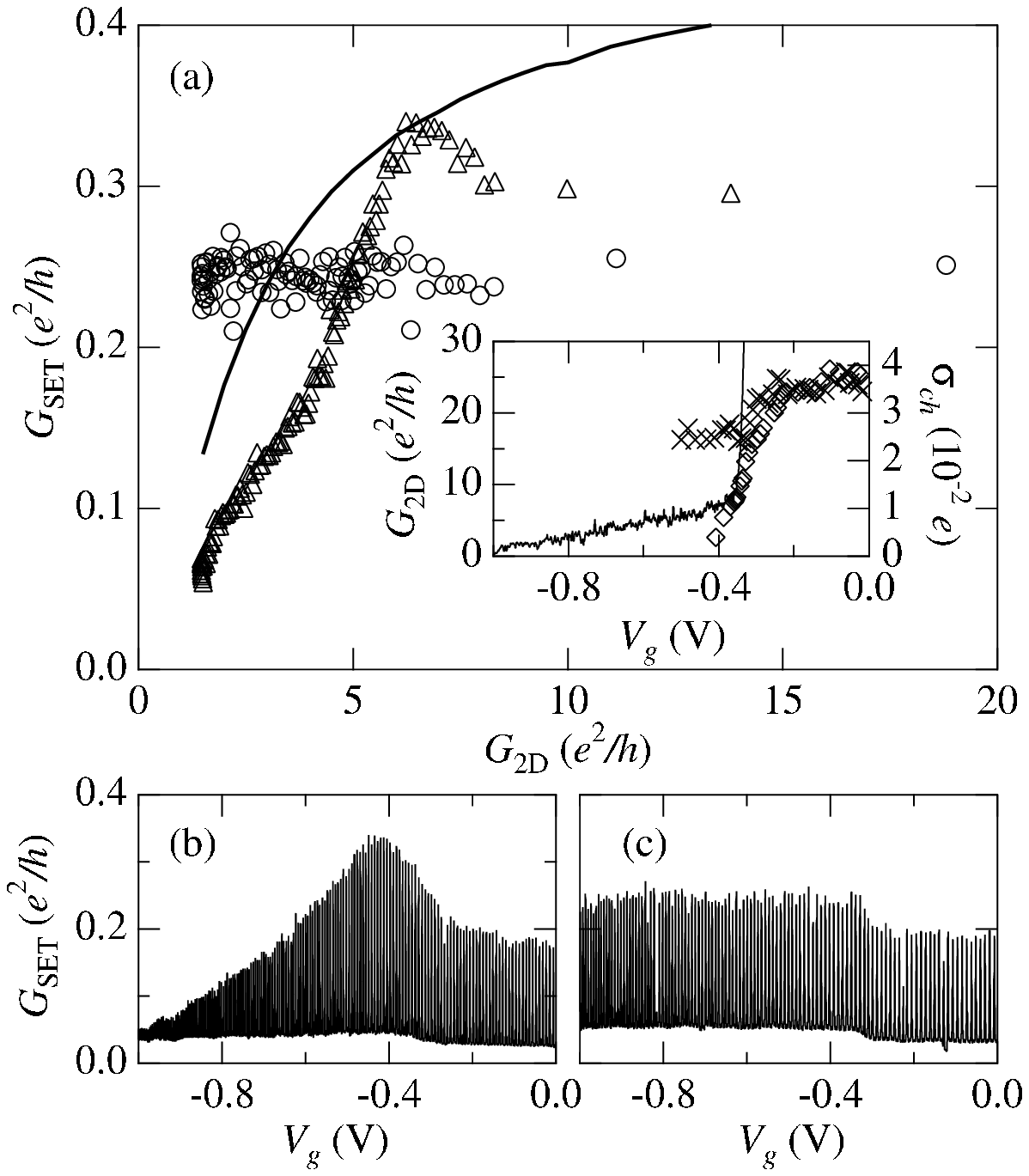} 
\caption{\label{gsetplot} (a) Maxima of \gset\ vs.\ \gtd\ in the stripe
($\circ$) and pool ($\triangle$) geometries.  Solid line:  \gsc vs.\
\gtd. Inset: \gtd\ (solid line) for the pool, and \sch\ in the stripe
($\times$) and pool ($\Diamond$) geometries vs.\ \vg.  (b) \gset\ vs.\
\vg\ in the pool geometry. (c) \gset\ vs.\ \vg\ in the stripe geometry.}
\end{figure}

Measurements were performed on two separate samples (S1 and S2) in a
dilution refrigerator in a four-probe voltage biased configuration; the
estimated electron temperature was \amount{100}{mK}.  High frequency
noise was excluded using standard techniques. A small capacitance
$\cg\approx\amount{20.3}{aF}$ (not shown in Fig.~\ref{sample}(c))
couples the six Au gates to the S-SET\@. The other sample parameters
such as the junction resistances $R_{1,2}$ and capacitances $C_{1,2}$,
the coupling capacitance \ctd\ and superconducting gap $\Delta$ are
given elsewhere \cite{Lu:2002}.  The charging energy $\ec=e^{2}/2\csig$
for sample S1 (S2) is \amount{118\: (77)}{\mu eV}  while the Josephson
energy $E_{J_{j}}=\frac{\rqu}{2R_{j}}\Delta$  averaged for the two
junctions is \amount{4.7\: (21.8)}{\mu eV}. Here $\rqu=\frac{h}{4e^2}$
is the superconducting resistance quantum and $\csig=\co+\ct+\ctd+\cg$.
We use standard lock-in techniques and voltage biases of 3 and
\amount{5}{\mu V\; rms} respectively to measure \gset\ and the
conductance \gtd\ across the series combination of the QPCs versus \vg\
in the pool and stripe geometries.  The results for S2 are shown in
Fig.~\ref{gsetplot}.

From Fig.~\ref{gsetplot}(b), we see that for the pool \gset\ rises by
nearly a factor of 2 as \vg\ becomes more negative, before dropping
rapidly.  Although \gtd\ vs. \vg\ is nearly identical in both cases,
\gset\ for the stripe rises only by $\sim$50\% and does not decrease,
even for the most negative \vg.  In both plots, $e$-periodic Coulomb
blockade oscillations are seen as the S-SET offset charge varies.  We
fit a smoothly varying function to the measured \gtd\ versus \vg\
(inset, Fig.~\ref{gsetplot}(a)) which we use to plot the maxima of
\gset\ versus \gtd\ in Fig.~\ref{gsetplot}(a).  The knee in \gtd\
at $\vg\approx\amount{-0.38}{V}$ corresponds to the appearance of
quantized plateaux in the individual QPC conductances.

To understand these results and the \iv\ curve in Fig.~\ref{sample} (b),
we begin with the rate of sequential
Cooper pair tunneling \cite{Averin:1990} through junction $j$, valid for
$E_{J}\ll\ec$ 
\begin{equation}\label{gameq}
\Gamma(\delta f^{(j)}) =
(\pi/2\hbar) E_{J_{j}}^{2}P(-\delta f^{(j)}) 
\end{equation} 
where $\delta f^{(j)} = f_{f} - f_{i} = (-1)^{j}4\ec(2N - n_{g}+1) -
2\alpha_{j}e V$ is the free energy difference for changing the number of
Cooper pairs $N$ by 1,
$\alpha_{j}=\textstyle{\frac{1}{2}}+(-1)^{j}(\co-\ct)/2\csig$, and
$n_{g} = \cg\vg/e$.  Here P(E) is the probability of exchanging energy
$E$ with the environment and can be expressed in terms of a correlation
function $K(t) = \rqu^{-1}\int_{-\infty}^{\infty}\frac{d\omega}{\omega}
\text{Re}[\zt(\omega)]\{\coth(\frac{\hbar \omega}{2\kb T})[\cos(\omega
t)-1] - i \sin(\omega t)\}$ via $P(E)=
\frac{1}{2\pi\hbar}\int_{-\infty}^{\infty}dt\;\exp[K(t)+i\frac{Et}{\hbar
} ]$ where $Z_{t}(\omega)$ is the total impedance seen by tunneling
electrons.

The general result for $\zt(\omega)$ within the model of
Fig.~\ref{sample}(c) is quite complex \cite{Ingold:1991}. In our case,
however, \zl\ (\ztd) dominates at low (high) frequencies and to an
excellent approximation
\begin{equation}\label{ztweq}
\text{Re}[\zt(\omega)]=%
\kappa_{1}\text{Re}[\zl(\omega)]+\frac{\kappa_{2}\ztd}{1+[\omega\ceff\kappa_{2}\ztd]^2}\label{zteq}
\end{equation}
where for junction 1(2)
$\kappa_{1}=\frac{(C_{2(1)}+\ctd)^2+C_{2(1)}^2}{\csig^2}$,
$\kappa_{2}=(\frac{\ctd}{\csig})^2$,
$\ceff=\frac{(C_{1}+C_{2})\csig}{\ctd}$ and we treat \ztd\ as a
resistance.  For \zl\ we begin with the impedance of a \textit{finite}
$RC$ line 
$Z_{RC}(\omega)=\sqrt{\frac{\rl}{i\omega\cl}}\tanh(\sqrt{i\omega\rl\cl
\ell^2} )$ where \rl\ and \cl\ are the resistance and capacitance per
unit length and $\ell$ is the total line length.  We are interested in
the long-time limit of $K(t)$ which is dominated by the low-frequency
part of $\zt(\omega)$.  In that limit,
$\zrc(\omega)\approx\frac{\rl\ell}{1+(\omega\rl\cl\ell^2/\sqrt{6})^2}$,
which we use for \zl\ in Eq.~\ref{zteq}.

A detailed analysis of $K(t)$ will be given elsewhere.  Here we note
that both parts of $\zt(\omega)$ in Eq.~\ref{zteq} have the same
form.   For \ztd, the corner frequency
$\omega_{c}=1/(\ceff\kappa_{2}\ztd)$ satisfies $\hbar\omega_{c}\gg\kb T$
and we may use the kernel of Ref.~\onlinecite{Wilhelm:2001}, while for
\zrc\ we find that $\omega_{c}=\sqrt{6}/\rl\cl\ell^2$ usually satisfies
$\hbar\omega_{c}\ll\kb T$ and requires different treatment. Since $K(t)$
is linear in $\text{Re}[\zt(\omega)]$, we may calculate $K(t)$ and
$P(E)$ separately for \ztd\ and \zl\ and find the total
$P_{\text{tot}}(E)$ as a convolution.  For \ztd, then, we have
$P_{2D}(E) = \frac{(2\pi\kappa_{2}\ceff\ztd\kb
T/\hbar)^{2\kappa_{2}/g}}{2\pi^2\kb
T}\text{Re}[e^{-i\pi\kappa_{2}/g}B(\frac{\kappa_{2}}{g} - \frac{i
E}{2\pi\kb T},1-\frac{2\kappa_{2}}{g})]$ where $g=\rqu/\ztd$ and
$B(x,y)$ is the beta function \cite{Wilhelm:2001}.  For \zrc, we find
that
\begin{eqnarray}\label{krc}
K_{RC}(t)=-\frac{2\kappa_{1}}{\grc}\bigg\{\pi\kb T %
|t|/\hbar+\frac{\pi}{2}[\cot(\frac{\hbar\grc}{2\kb T \tau})\nonumber\\
 \mbox{}-i \text{sign}(t)](e^{-|t|/\tau}-1)\bigg\}
\end{eqnarray}
is valid for $\frac{\hbar\grc}{2\kb T \tau}\alt\pi/2$, where
$\grc=\rqu/(\rl\ell)$ and $\tau=\rqu\cl\ell/\sqrt{6}$.  From this we
calculate
\begin{eqnarray}\label{prc}
P_{RC}(E) = \frac{\tau}{\pi\grc\hbar}e^{\gamma_{3}(\grc)}%
\text{Re}\big[e^{-i\frac{\kappa_{1}\pi}{\grc}}\gamma_{2}(\grc)^{-\gamma_{1}(\grc)}\nonumber\\
\times\left\{\Gamma\left[\gamma_{1}(\grc)\right]%
-\Gamma\left[\gamma_{1}(\grc),\gamma_{2}(\grc) \right]\right\}\big]
\end{eqnarray}
where $\gamma_{1}(g)=\frac{2\pi\kappa_{1}\kb T \tau}{\hbar g^2}
-iE\tau/\hbar g$, $\gamma_{2}(g)=\gamma_{3}(g)- i\pi\kappa_{1}/g$,
$\gamma_{3}(g)=\frac{\kappa_{1}\pi}{g}\cot\left(\frac{\hbar  
g}{2\kb T\tau}\right)$ and $\Gamma(x,y)$ is the incomplete gamma
function.

\begin{figure}[t]
\includegraphics[width=2.9in]{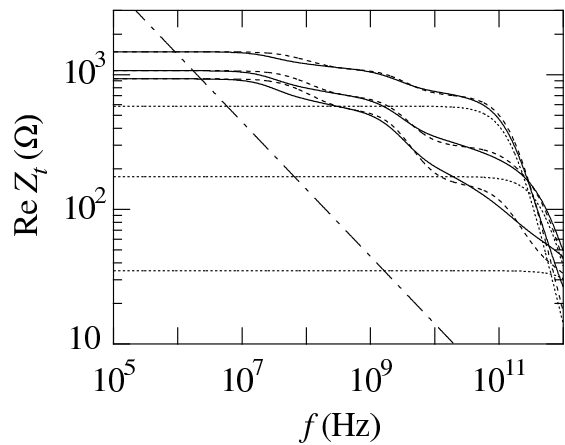}
\caption{\label{ztw} Calculated $\zt(\omega)$ for (bottom to top)
$\ztd=258$, 1291 and \amount{4302}{\Omega} for three different models: cascaded
$RC$ lines (solid), approximate series combination (dashed) and 
\ztd\ only (dotted).  The dash-dotted line is the impedance
of an infinite line with $\rl=\amount{2.9}{M\Omega/m}$ and
$\cl=\amount{23}{nH/m}$. }
\end{figure}

To proceed we need an accurate model of $\zt(\omega)$. For the $\vg=0$
\iv\ curve to be linear at $V\approx\amount{8}{\mu V}$, $\zt/\rqu$ must
be nonnegligible at frequencies of order $e V/h \approx\amount{2}{GHz}$.
Since then $\ztd\approx\rsq=\amount{20}{\Omega}$, \zl\ must dominate
\zt\ for small \ztd.  We therefore consider the structure of our leads,
which vary in width $w$ from \amount{100}{nm} to \amount{20}{\mu m}. The
\amount{100}{nm} section has length $\ell=\amount{1}{\mu m}$,
contributes only \amount{50}{\Omega} to $\zl(0)$, and is not considered
further.  For the remaining sections with $w=$ 0.4, 1.0, 10 and
\amount{20}{\mu m}, and  $\ell=$ 9, 57, 253 and \amount{375}{\mu m} we
use $\rl\approx\rsq/(w+5.8h)$ and
$\cl\approx\varepsilon\varepsilon_{0}(w/h+1.393)$ where
$h=\amount{50}{nm}$ is the 2DEG depth and $\varepsilon=13$ the
dielectric constant of GaAs to calculate $\rl=$ 29, 16, 1.9 and
\amount{1.0}{M\Omega/m} and $\cl=$ 1.1, 2.5, 23 and \amount{46}{nF/m}.
These four sections form a cascaded $RC$ line which determines
$\zl(\omega)$.  The total $\zt(\omega)$ calculated from Eq.~\ref{ztweq}
is shown for different values of $\ztd=1/(4\gtd)$ for the pool geometry
in Fig.~\ref{ztw}.  The cascaded form for $\zl(\omega)$ is quite
complex.  For our calculations we take
$\text{Re}[\zl(\omega)]=\sum_{i}\text{Re}[\zrc^{(i)}]$ where the
$\zrc^{(i)}$ are the impedances of the individual sections, a very good
approximation to the more exact result, as shown.  Note that this model
predicts  a significant \zt\ at \amount{2}{G Hz} dominated by \zl\
(\ztd) for small (large) \ztd.  For the stripe, \ztd\ approaches
$\rqpc/2$ at zero frequency and the much lower stripe resistance
$\rstr\approx\amount{200}{\Omega}$ at frequencies above $1/\rqpc\cstr$
where $\cstr\approx\amount{0.3}{pF}$ is its capacitance to ground. At
high frequencies, then, \zt\ in the stripe is always dominated by \zl,
even for large negative \vg. We have also shown the impedance for \ztd\
alone, and for an infinite $RC$ line with $w=\amount{10}{\mu m}$ and
\rl\ chosen to give the correct $\zl(0)$ if the line were finite.  The
latter two models give a small \zt\ for small \ztd\ at the relevant
frequencies, and cannot explain the linear region in our $\vg=0$ \iv\
characteristics.

Using $P_{2D}(E)$ and $P_{RC}(E)$ above, we numerically convolve the
$P_{i}^{(j)}(E)$ for junction $j$ and section $i$ to find
$P_{\ell}^{(j)}(E) = P_{1}^{(j)}(E)\ast P_{2}^{(j)}(E)\ast
P_{3}^{(j)}(E)\ast P_{4}^{(j)}(E)$.  We then calculate
$P_{\text{tot}}^{(j)}(E)=P_{\ell}^{(j)}(E)\ast P_{2D}(E)$ for different
\gtd\ and set up a master equation using the rates in Eq.~\ref{gameq} to
calculate the S-SET current and conductance \gsc.  The results for \gsc\
in the pool geometry are shown as the solid line in
Fig.~\ref{gsetplot}(a); we scale \gsc\ to match the maximum measured
value $\gset^{\text{max}}\approx 0.34G_{0}$ at $\gtd^{\text{max}}\approx
6.5G_{0}$ but use no other variable parameters.  \gsc\ agrees reasonably
well with \gset\ for $\gtd<\gtd^{\text{max}}$ although it rises less
steeply with \gtd.  In this regime $P_{2D}(E)$ is broad and inelastic
transitions suppress the coherent supercurrent.  For
$\gtd>\gtd^{\text{max}}$, \gsc\ gradually saturates at $0.47 G_{0}$. 
For $\gtd\gg\gtd^{\text{max}}$, $P_{2D}(E)\approx \delta(E)$ ({\itshape
i.\ e.}, only elastic transitions are likely) and $P_{\ell}(E)$
dominates the \iv\ characteristic.  No nonmonotonic behavior occurs in
\gsc, in agreement with Wilhelm \etal\ The drop in \gset\ for
$\gtd>\gtd^{\text{max}}$ must arise from  other physics.

\begin{figure}
\includegraphics[width=2.9in]{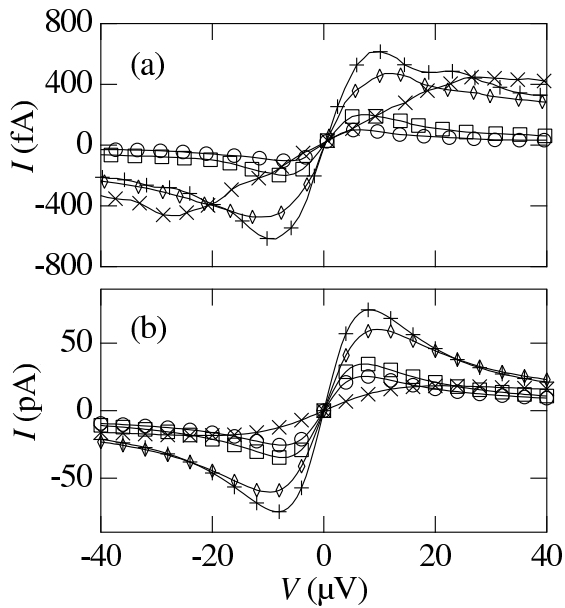}
\caption{\label{ivplot}Measured (a) and calculated (b) \iv\
characteristics for an unconfined 2DEG ($\circ$),  $\vg =
\amount{-0.3}{V}$ ($\Box$), and $\ztd = 1613$ (+),  2151 ($\diamond$)
and \amount{6453}{\Omega} ($\times$). To fit the data at $\vg = 0$ and
\amount{-0.3}{V}, we use $\sigma_{ch} = 0.07$ and $0.05 e$,
respectively.  For the remaining curves we take $\sigma_{ch} = 0$. }
\end{figure}

We can model the drop by assuming that the measured $I$ and \gset\ are
actually averaged over charge states close to $n_{g}$, due to charge
motion in the substrate \cite{Eiles:1994}. We expect charge averaging to
be most pronounced when the 2DEG is unconfined, and least for small
\gtd.  Assuming a Gaussian distribution of charge states with mean $n_g$
and variance $\sigma_{ch}$, we calculate the average
$\langle\gsc\rangle$. We do not know the absolute size of $\sigma_{ch}$,
so we assume for the pool that $\sigma_{ch}=0$ for
$\gtd<\gtd^{\text{max}}$. For $\gtd>\gtd^{\text{max}}$ we find the
values of $\sigma_{ch}$ which give $\langle\gsc\rangle=\gset$ for the
pool and stripe and plot the results in the inset to
Fig.~\ref{gsetplot}(a).  In both cases $\sigma_{ch}$ is just below $0.04
e$ near $\vg=\amount{0}{V}$ and  drops near $\vg =\amount{-0.28}{V}$.
For the stripe, $\sigma_{ch}$ saturates at just above $0.02 e$, about
half the drop for the pool.  The model seems reasonable given the small
$\sigma_{ch}$ required to explain the discrepancies with the
environmental theory.

We gain further confidence in the model by comparing measured and
calculated \iv\ characteristics for the pool, as shown in Fig.~\ref{ivplot} for S1. 
For increasing confinement $I$ first rises at all voltages,
with little or no broadening of the linear region ($\vg = 0$ and
\amount{-0.3}{V} and $\ztd=\amount{1613}{\Omega}$).  This corresponds to
a reduction in $\sigma_{ch}$ with little change in $P_{\text{tot}}(E)$;
inelastic transitions in the leads dominate the energy exchange. 
Eventually, $\sigma_{ch}=0$ and \ztd\ is large enough to
affect $P_{\text{tot}}(E)$, causing  $I$ to decrease (especially at low bias)
and the linear region to broaden.  The level of agreement between the
shape and evolution of the measured and calculated curves is
surprisingly good, given the uncertainties involved.  While the
calculated current is much larger than is measured, such discrepancies
are common in small tunnel junction systems \cite{Eiles:1994}.

In conclusion, we have measured the effects of dissipation on transport
in an S-SET for which the environment can be varied locally. We find
good agreement with a model in which fluctuations in the leads and
low-frequency switching between charge states dominate for low
confinement (large \gtd), while for strong confinement (small \gtd)
fluctuations coupled via the capacitance \ctd\ dominate.  The model
accounts well for the evolution of \gset\ and the \iv\ curves as \gtd\
is varied.  We believe a convolved $P_{\text{tot}}=P_{\ell}\ast P_{2D}$
is likely required to interpret the results of the Berkeley group, which
may explain the discrepancies between their results and the scaling
theory of Wilhelm, \etal

This research was supported at Rice by the NSF under Grant No.\
DMR-9974365 and by the Robert A. Welch foundation, and at UCSB by the
QUEST NSF Science and Technology Center.  One of us (A. J. R.)
acknowledges support from the Alfred P. Sloan Foundation.  We thank A.
C. Gossard for providing the 2DEG material.


\end{document}